\documentclass{article}
\usepackage{graphics}
\usepackage{amsmath}
\usepackage{amsfonts}
\usepackage{amssymb}
\usepackage{epsfig}

\def\ll{{\langle}}
\def\rr{{\rangle}}

\begin{document}

\title{{Group Theoretical Formulation of Quantum Partial Search Algorithm}}

\author{
{\Large  Vladimir E. Korepin  and Brenno C. Vallilo }\\[3mm]
{\it C.N. Yang Institute for Theoretical Physics}\\
{\it SUNY, Stony Brook, NY 11794-3840, USA}
}

\date{June 2006}
\maketitle

\begin{abstract}

Searching and sorting  used  as a subroutine in many important 
algorithms. 
Quantum algorithm can find a target item in a database faster than any
classical algorithm. One can trade accuracy for speed and find a part of the
database (a block) containing the target item even  faster, this is  partial
search. An example is the following: exact address of the target item is given by a sequence
 of many bits, but we need to know only some of them.
More generally  partial search considers the following problem: a  database is
 separated into    several { blocks}.
We  want to find a block with the target item, not the target  item itself.
In this paper we reformulate  quantum partial search algorithm  in terms of 
group theory.

\end{abstract}

\section{Introduction}
\label{intro}
Database search is used as a subroutine in may important algorithms 
\cite{clrs,theorem,cleve}.
Grover discovered a quantum algorithm that searches faster than any classical
algorithm \cite{Grover}.  It consists of repetition of the Grover iteration
$\hat{G}_1$. We shall call it global iteration, see (\ref{iter}).
The number of  repetitions [queries to the oracle] is:
\begin{equation}
j_{\mbox{full}}= \frac{\pi}{4} \sqrt{N} \qquad \mbox{as} \qquad N\rightarrow
\infty \label{full}
\end{equation}
for a database with  $N$ entries.
There is no faster quantum algorithm 
\cite{Bennett,Boyer,zalka}.
Nevertheless if we need less information then the search can be performed 
faster. For example if the exact address of the target item in a database is
given by a sequence of many bits $b_1 b_2 \ldots b_n$, but we want to know only three  of them, we can do it faster then (\ref{full}).
This is an example of partial search.
A partial search considers the following problem: a  database is
 separated into    $K$ { blocks} of  a  size $b={N}/{K}$.
We  want to find a block with the target item, not the target  item itself.
The   block with the target item is  called the target block, all other blocks are non-target blocks.
Grover and   Radhakrishnan suggested a  quantum algorithm for a partial
search in \cite{jaik}.
Partial search naturally arise in list  matching
 \cite{hei}.
 Classical partial search takes
  $\sim (N-b)$ queries, but
quantum  algorithm takes only 
$\sim ( \sqrt{N}- \mbox{coeff}\sqrt{b}) $ queries. Here $\mbox{coeff} $ is a positive number, which has a limit, when number of items in each block is very large $b\rightarrow \infty$.
Grover-Radhakrishnan algorithm uses several global iterations $\hat{G}^{j_1}_1$ and then several 
local  iteration $\hat{G}^{j_2}_2$, see (\ref{liter}).
 Grover-Radhakrishnan algorithm was  simplified and clarified in \cite{kg} and  optimized   in
 \cite{kor}
 [the number of  queries to the oracle was minimized, positive $\mbox{coeff}$ was increased]. Other partial search algorithms were studied in \cite{liao}.
 The algorithm for  blocks of finite size was formulated in \cite{choi}. Partial search algorithm for a database with multiple target items was formulated  in \cite{ck}.

\section{The GRK Algorithm for Partial Search}

To introduce the partial search algorithm it is useful to first
remind the full Grover search. We shall consider a database with one
target item. The Grover algorithm finds the target state
$|t\rangle$ among an unordered set of $N$ states, which is called the
database. In the classical case, the items in the database are labeled  by a 
 sequence of bits. In the quantum case  these sequences label orthonormal
 basis in a linear  space  $|x\rr$. The search is performed by repeating global iteration
which is defined in terms of two operators. The first changes the
sign of the target state $|t\rangle$ only:
\begin{equation}
\hat{I_t}=\hat{I}-2|t\rangle \langle t|, \qquad \langle
t|t\rangle=1, \label{target}
\end{equation}
where $\hat{I}$ is the identity operator and $|t\rr\ll t|$ projects
on the target item. The second operator,
\begin{equation}
\hat{I}_{s_1}=\hat{I}-2|s_1\rangle \langle s_1|, \label{average}
\end{equation}
changes the sign of the uniform superposition of all basis states
$|s_1\rangle$
\begin{equation}
|s_1\rangle = \frac{1}{\sqrt{N}}\sum_{x=0}^{N-1}|x\rangle , \qquad
\langle s_1|s_1\rangle =1 . \label{ave}
\end{equation}
The {\bf global} iteration is defined as a unitary operator
\begin{equation}
\hat{G_1}=-\hat{I}_{s_1}\hat{I}_t . \label{iter}
\end{equation}

To describe  partial search we need to introduce   a {\bf local} search.
Local search is a search inside of each block done simultaneously in all 
blocks.  In one block local search   acts as
\begin{equation}
\hat G_2=-\hat I_{s_2}\hat I_t, \label{liter}
\end{equation}
where $\hat I_t$ is the same operator appearing in the global search (\ref{target}), but
$\hat I_{s_2}$ acts in each block as
\begin{equation}\label{is2}
\hat{I}_{s_2}{\big |}_{block}=\hat{I}{\big |}_{block}-2|s_2\rangle
\langle s_2|, \qquad |s_2\rangle =
\frac{1}{\sqrt{b}}\sum_{\mbox{\scriptsize{one block}}}|x\rangle.
\end{equation}
Local search in  the whole database is the direct sum of
$\hat I_{s_2}$ with respect to all blocks.

Let us mention  eigenvectors and eigenvalues of global iteration, see \cite{Brass}:
\begin{eqnarray}
\hat{G_1}|\psi^{\pm}_1\rangle  = \lambda^{\pm}_1
|\psi^{\pm}_1\rangle , \qquad \lambda^{\pm}_1 =\exp[{\pm 2i
\theta_1}] , \qquad \\
 |\psi^{\pm}_1\rangle  = \frac{1}{\sqrt{2}}|t\rangle \pm \frac{i}
{\sqrt{2}}\left(
\sum^{N-1}_{\stackrel{\mbox{\small{x=0}}}{\mbox{\small{x  $\neq$
t}}}} \frac{|x\rangle}{\sqrt{(N-1)}} \right), \label{value}
\end{eqnarray}
where $\theta_1$ is
$$\sin^2\theta_1=\frac{1}{N}$$

In the case of local search  $\hat G_2$ they are given by
\begin{eqnarray}
\hat{G}_2|\psi^{\pm}_2\rangle =\lambda^{\pm}_2
|\psi^{\pm}_2\rangle, \qquad
 \lambda^{\pm}_2  = \exp[{\pm 2i \theta_2}] , \label{vector2}  \qquad
|\psi^{\pm}_2\rangle   =\frac{1}{\sqrt{2}}|t\rangle \pm
\frac{i}{\sqrt{2}} |\mbox{ntt}\rangle
\end{eqnarray}
Amplitudes of all non-target items in the target blocks are the same.
So we can only follow the amplitude of   $|\mbox{ntt}\rangle $ which is the
normalized sum of all non-target items in the target block:
\begin{equation}
|\mbox{ntt}\rangle = \frac{1}{\sqrt{b-1}} \sum_{\stackrel{x \neq
t}{\mbox{\tiny{target block}}}} |x\rangle,\qquad \langle
\mbox{ntt} |\mbox{ntt}\rangle =1, \quad \langle \mbox{ntt} |\mbox{t}\rangle =0. \label{vntt}
\end{equation}
Here the angle $\theta_2$ is given by
\begin{equation}
\sin^2 \theta_2 =\frac{K}{N}=\frac{1}{b}. \label{ang2}
\end{equation}

\subsection{Three Dimensional Space}

Amplitudes of all items in non-target blocks are the same.
So we can only follow the amplitude of  $|u\rangle$.
 The vector
$|u\rangle$ is given by
\begin{eqnarray}
|u\rangle = \frac{1}{\sqrt{b(K-1)}} \sum_{\stackrel{\mbox{all
items in all}}{\mbox{{non-target blocks}}}} |x\rangle , \qquad\quad
\langle u|u\rangle =1 .\label{ntb}
\end{eqnarray}
Together with the target item $|t\rr$ and $|ntt\rr$ [see (\ref{vntt})] the unite vector  $|u\rangle$ form a orthonormal  basis in three dimensional linear space: $\ll t|u\rr=\ll ntt|u\rr=0$.

For example the eigenvectors of global iterations $\hat G_1$ in Eq. (8)
can be written as
\begin{equation}|\psi^{\pm}_1\rangle =
\frac{1}{\sqrt{2}}|t\rangle \pm \frac{i}
{\sqrt{2}}\left(
\sum_{\stackrel{\mbox{\small{x=0}}}{\mbox{\small{x  $\neq$
t}}}}^{N-1} \frac{|x\rangle}{\sqrt{(N-1)}} \right)= \end{equation}
$$=\frac{1}{\sqrt{2}}|t\rangle \pm \frac{i}
{\sqrt{2}}\left( \sqrt{\frac{b-1}{N-1}}|ntt\rangle +
\sqrt{\frac{b(K-1)}{N-1}}|u\rangle \right).
$$

Below we shall describe the partial search algorithm as a three dimensional 
matrix (\ref{grk}).

Meanwhile let us remind 
the GRK algorithm for partial search. The partial search of \cite{kor} creates a vector
\begin{equation}
|d\rangle =\hat{G}_1 \hat{G}^{j_2}_2 \hat{G}^{j_1}_1|s_1\rangle
.\label{gl}
\end{equation}

Note that this algorithm uses a sequence of global-local operators.
The final operation $\hat G_1$ is necessary since $\hat G_2$ acts
trivially on $|u\rr$, i.e. $\ll u|\hat G_2=\ll u|$. The final state
$|d\rangle $ should have zero amplitudes of each item in non-target
blocks, in other words it should satisfy
$$\langle u|d\rangle =0,$$
which means that a measurement will reveal the position of the
target block. We consider large blocks $b=N/K\rightarrow \infty$. The
number of blocks $K$ is a finite number ($K=2,3,\dots$), it is an important parameter in the algorithm. It is useful to
introduce an angle $\gamma$ defined by
$$\sin(\gamma)=\frac{1}{\sqrt{K}},\quad 0<\gamma\leq \frac{\pi}{4}.$$

In the limit $b\to\infty$ it was shown in \cite{jaik} that
the total number of iterations grows as
$$j_1=\frac{\pi}{4}\sqrt{N}-\eta_K\sqrt{\frac{N}{K}},\quad
j_2=\alpha_K\sqrt{\frac{N}{K}},$$
where the coefficients $\eta_K$ and $\alpha_K$ have a well defined limit.
The total number of queries to the oracle is $j_1+j_2+1\to
{\pi}/{4}\sqrt{N}-(\eta_K-\alpha_K)\sqrt{{N}/{K}}$.

One of the authors found the optimal values of
$(\eta_K,\alpha_K)$ (see \cite{kor}) such that total number of queries to the oracle  is minimal:
$$\tan{\left(  \frac{2\eta_K}{\sqrt{K}} \right) } =\frac{ \sqrt{3K-4}}{K-2},\quad
\cos{(2\alpha_K)}=\frac{K-2}{2(K-1)}$$

\section{$O(3)$ Group Formulation of GRK}
\label{othree}

The GRK  is the fastest among partial search algorithms, which
 use  the sequence
$\hat{G}_1 \hat{G}^{j_2}_2 \hat{G}^{j_1}_1$. Partial  search algorithms
using some other sequences were considered in \cite{liao}, but no acceleration 
was found.
 In order to prove that the GRK algorithm is the
fastest among all partial search algorithms, which use other sequences of  $\hat G_1$ and
$\hat G_2$
one has to prove the following:\\
{\bf Conjecture:} Start from an arbitrary
vector $|\phi\rangle$ in the three dimensional space with the basis
$(|t\rr,|ntt\rr ,|u\rr )$. If the sequence
of local-global-local-global
operations can find the target block
$$\langle u| \hat{G}^{j_4}_1 \hat{G}^{j_3}_2
\hat{G}^{j_2}_1\hat{G}^{j_1}_2|\phi\rangle=0,$$
then there exists a global-local-global sequence
such that
$$ \langle u| \hat{G}^{\tilde j_3}_1 \hat{G}^{\tilde j_2}_2
\hat{G}^{\tilde j_1}_1|\phi\rangle=0,$$
it finds the target black faster, i.e.
$\tilde j_3+\tilde j_2+\tilde j_1\leq j_1+j_2+j_3+j_4$.$\qquad \maltese$ \\

If the conjecture is true then  GRK is the fastest among partial search 
algorithms using arbitrary sequence of local and global searches.

{\bf Proof:} Assume a general algorithm constructed with global $\hat G_1$
and local $\hat G_2$ iterations with $n$ letters given by
\begin{equation}\label{seqgrk}\ll u| \hat G_1^{j_n}\hat G_2^{j_{n-1}}\hat G_1^{j_{n-2}}
\hat G_2^{j_{n-3}} \hat G_1^{j_{n-4}} \hat G_2^{j_{n-5}} \cdots \hat
G_1^{j_3}\hat G_2^{j_2}\hat G_1^{j_1}|s_1\rr=0,
\end{equation}
where $j_i\geq 0$ for $i\leq n-1$ and $j_n>0$. In particular we can
have $j_1=0$, so the sequence can also start with $\hat G_2$. The final operation has to be $\hat G^{j_n}_1$ since $\hat G_2$ acts
trivially on $|u\rr$, i.e. $\ll u|\hat G_2=\ll u|$.
Equation (\ref{seqgrk}) can be written as
$$\ll u|\hat G_1^{j_n}\hat G_2^{j_{n-1}}\hat G_1^{j_{n-2}}
\hat G_2^{j_{n-3}} |\phi \rr= 0,$$
where
$|\phi\rr=\hat{G}^{j_{n-4}}_1\cdots \hat{G}^{j_2}_2\hat{G}^{j_1}_1|s_1\rr$
Using the conjecture, this can be reduced to
$$\ll u| \hat{G}^{\tilde j_{n-1}}_1 \hat{G}^{\tilde j_{n-2}}_2
\hat{G}^{\tilde j_{n-3}}_1|\phi\rangle=0,$$
where $\tilde j_{n-1}+\tilde j_{n-2}+\tilde j_{n-3}\leq
j_n+j_{n-1}+j_{n-2}+j_{n-4}$.
We now have another sequence
$$\ll u| \hat{G}^{\tilde j_{n-1}}_1 \hat{G}^{\tilde j_{n-2}}_2
\hat{G}^{\tilde j_{n-3}}_1\hat{G}^{j_{n-4}}_1 \hat G_2^{j_{n-5}}
|\phi^\prime\rangle=0,$$ where $|\phi\rr=\hat{G}^{j_{n-4}}_1\hat
G_2^{j_{n-5}} |\phi^\prime\rr$. 
We can put two powers of $\hat{G}_1$ together
$$\ll u| \hat{G}^{\tilde j_{n-1}}_1 \hat{G}^{\tilde j_{n-2}}_2
\hat{G}^{\tilde j_{n-3}+j_{n-4}}_1 \hat G_2^{j_{n-5}}
|\phi^\prime\rangle=0,$$ 
and use the  conjecture  again,
starting with $|\phi'\rr$. After several iterations, (\ref{seqgrk})
will be reduced to
\begin{equation}\label{final}\ll u|\hat{G}^{k_4}_1\hat{G}^{k_3}_1
\hat{G}^{k_2}_2\hat{G}^{k_1}_1|s_1\rr=0.\end{equation}

It was proved  in \cite{liao} that the GRK algorithm is the fastest
among all possible algorithms in the form (\ref{final}).
 This means that in order to proof that  GRK algorithm is the fastest among all partial search algorithms consisting of arbitrary sequence of of $\hat{G}_1$ and $\hat{G}_2$
it is enough to prove  the conjecture.$\qquad \heartsuit$\\

$\bullet$ Let us reduce partial search to $O(3)$ group.

In subsection 2.1 we explained that the partial search algorithm acts 
naturally in three dimensional space with the orthonormal basis:
 target item $|t\rr$, non-target items in the target block $|ntt\rr$ and all 
items in non-target blocks $|u\rr$.
 Search operations are
rotations in three dimensional space spanned by these three vectors.
All the  vectors involved in present quantum search problem can
be written in this basis as
\begin{equation}\label{vector} |V\rr=\left(\begin{array}{c}a\\b\\c\end{array}\right),
\end{equation}

In the above equation $(a,b,c)$ are
the real coefficients in the base defined by $(|t\rr,|ntt\rr,|u\rr)$, meaning
\begin{equation}\label{vectorform}|V\rr=a|t\rr+b|ntt\rr+c|u\rr
\end{equation}
For example, the  initial state (\ref{ave}) can be written as:
\begin{equation}
|s_1\rangle =\left(\begin{array}{c} \sin \gamma \sin \theta_2\\
\sin \gamma \cos \theta_2 \\ \cos \gamma \end{array}\right) \label{initial}
\end{equation}
and the local uniform state (\ref{is2}) is
\begin{equation}
|s_2\rangle =\left(\begin{array}{c} \sin \theta_2 \\
\cos \theta_2 \\ 0\end{array}\right) \label{local
initial}
\end{equation}

From this basic relations and the definitions of $\hat G_2$ we can calculate
its $j_2$ power of local search:
\begin{equation}
\hat{G}^{j_2}_2=\left( \begin{array}{clcr}
\cos (2j_2\theta_2) & \sin (2j_2\theta_2) &0 \\
-\sin (2j_2\theta_2) & \cos (2j_2\theta_2) &0 \\
0 & 0& 1
\end{array}
\right) \label{loc}
\end{equation}
The ordering of eigenvectors is  $|t\rangle$, $|ntt\rangle $  and
$|u\rangle $. The matrix  has three eigenvectors:
\begin{eqnarray}
\hat{G}^{j_2}_2|v_2^{\pm}\rangle =\exp (\pm 2i\theta_2
j_2)|v_2^{\pm}\rangle ,\quad \hat{G}^{j_2}_2|v_2^{0}\rangle
=|v_2^{0}\rangle
\end{eqnarray}
Where the eigenvectors can be written as
\begin{equation}
|v_2^{\pm}\rangle=\frac{1}{\sqrt{2}}\left(\begin{array}{c}
1\\
\pm i \\
0
\end{array} \right), \qquad |v_2^{0}\rangle= \left(\begin{array}{c}
0\\
0 \\
1
\end{array} \right).
\end{equation}

In the same way,  $j_1$ repetitions of
the global iterations  (\ref{iter}) is \footnote{here we use
$c(\cdot)=\cos(\cdot)$ and $s(\cdot)=\sin(\cdot)$}

\begin{equation}
\hat{G}^{j_1}_1= \label{mat}
\end{equation}
\\
\begin{equation*}\left( \begin{array}{ccc} c(2j_1\theta_1), &
s(2j_1\theta_1) s(\gamma), &
s(2j_1\theta_1) c(\gamma) \\
-s(2j_1\theta_1)s(\gamma),\quad  & (-1)^{j_1}c^2 (\gamma) +c (2j_1\theta_1) s^2 (\gamma),\quad & s(\gamma) c(\gamma) \left( (-1)^{j_1+1}+c(2j_1\theta_1)\right) \\
- s(2j_1\theta_1) c(\gamma), \quad &s(\gamma) c(\gamma)
\left( (-1)^{j_1+1}+c(2j_1\theta_1)\right),  &
(-1)^{j_1}s^2(\gamma) +c(2j_1\theta_1) c^2 (\gamma)
\end{array}
\right) 
\end{equation*}
This is a simplified asymptotic expression valid in the limit of
large blocks $b\rightarrow \infty$. The
matrix  has three eigenvectors
\begin{eqnarray}\label{reflection}
\hat{G}^{j_1}_1|v_1^{\pm}\rangle =\exp (\pm 2i\theta_1
j_1)|v_1^{\pm}\rangle ,\quad \hat{G}^{j_1}_1|v_1^{0}\rangle
=(-1)^{j_1}|v_1^{0}\rangle
\end{eqnarray}
where the eigenvectors are
\begin{equation}
|v_1^{\pm}\rangle=\frac{1}{\sqrt{2}}\left(\begin{array}{c}
1\\
\pm i\sin \gamma \\
\pm i\cos \gamma
\end{array} \right), \qquad |v_1^{0}\rangle= \left(\begin{array}{c}
0\\
\cos \gamma \\
-\sin \gamma
\end{array} \right).
\end{equation}

 $\maltese \qquad $ It is also possible to represent  the whole  GRK algorithm as a  matrix

\begin{equation}  \hat G_{GRK}=\hat{G}_1\hat{G}^{j_2}\hat{G}^{j_1}_1= \end{equation}
\begin{equation} \label{grk}
\left( \begin{array}{ccc} 0 & \frac{1}{2\sqrt{K-1}}-
\frac{1}{2}\sqrt{\frac{3K-4}{K}} & \frac{1}{2}+
\frac{1}{2}\sqrt{\frac{3K-4}{K(K-1)}} \\
0 &  \frac{1}{2}+ \frac{1}{2}\sqrt{\frac{3K-4}{K(K-1)}} &
-\frac{1}{2\sqrt{K-1}}+
\frac{1}{2}\sqrt{\frac{3K-4}{K}} \\
-1 & 0 & 0 \end{array}\right),\end{equation} 
It has the
form,
\begin{equation}
\left( \begin{array}{ccc}0\quad&a\quad&b\quad \\ 0\quad &b\quad
&-a\quad \\-1\quad&0\quad&0\quad \end{array}\right),
\end{equation}
where $a$ and $b$ satisfy $a^2+b^2=1$ which shows that the GRK matrix
is an element of the group $O(3)$ $\qquad \qquad \heartsuit$.

\subsection{Reformulation in terms of $SO(3)$ Group}

We see in (\ref{reflection}) that the matrix corresponding to the
operator $\hat G_1$ has a {\it reflection} if $j_1$ is odd. This fact
makes the analysis of the algorithm difficult, since algorithms with
even and odd powers of $\hat G_1$ have a different behavior
\cite{liao}. To overcome this problem, we can reformulate the  algorithm
in such a way that it will use 
only even powers of $\hat G_1$. To do that we have to  introduce auxiliary search defined by
\begin{equation}\label{ga}\hat G_a^j=\hat G_1 \hat G_2^j \hat G_1.
\end{equation}
Now it is necessary to show how the introduction of this new operator is done
inside the algorithm. We consider a general sequence (\ref{seqgrk}) of
$\hat{G}_1$ and $\hat{G}_2$

\begin{equation}\label{generaltwo}
 \ll u| \hat G_1^{j_n}\hat G_2^{j_{n-1}}\hat G_1^{j_{n-2}}
\hat G_2^{j_{n-3}} \cdots \hat G_1^{j_3}\hat G_2^{j_2}\hat G_1^{j_1}|s_1\rr=0,
\end{equation}
We can always make the total number of  $\hat{G}_1$ factors  
($j_n+j_{n-2}+\cdots +j_3+j_1$)
to be an even number. We can  add one extra factor  $\hat{G}_1$ in the beginning  using the fact that $\hat{G}_1|s_1\rr=|s_1\rr +O({1}/{\sqrt{b}}) $:

\begin{equation}
 \ll u| \hat G_1^{j_n}\hat G_2^{j_{n-1}}\hat G_1^{j_{n-2}}
\hat G_2^{j_{n-3}} \cdots \hat G_1^{j_3}\hat G_2^{j_2}\hat G_1^{j_1}
\hat{G}_1|s_1\rr=0,
\end{equation}

 We can now consider only
sequences with even total number  of $\hat{G}_1$ factors.
Individual powers of  $\hat{G}_1$ can still be odd.
Each odd power of  $\hat{G}_1$ we represent as even multiplied by on
 $\hat{G}_1$ factor. Since the total number of  $\hat{G}_1$ factors is even,
single   $\hat{G}_1$ can only occurs in pairs.
This means that in the string of operators we can chose single
$\hat{G}_1$'s such that we have
\begin{equation}\label{changega}\ll u| \cdots \hat{G}_1 \hat{G}^{k_1}_2
\cdots \hat{G}^{k_2}_2\hat{G}_1 \cdots |s_1\rr=0,\end{equation} where
between the two factors of $\hat{G}^{k_1}_2$ and $\hat{G}^{k_2}_2$ we
have only  even  powers of $\hat{G}_1$'s. Now one uses the definition of
$\hat{G}_a$ given in Eq. \ref{ga} and rewrites Eq. \ref{changega} as
$$ \ll u| \cdots \hat{G}^{k_1}_a \cdots \hat{G}^{k_2}_a \cdots |s_1\rr=0.$$
Here we used $\hat{G}_1^2=1+ O(1/\sqrt{b  })$.

A general algorithm will now be a sequence of the three operators global search $\hat{G}_1$ raised in even powers, local search 
$\hat{G}_2$ and auxiliary search $\hat{G}_a$:

\begin{equation}
\ll u| \hat{G}^{j_n}_1 \hat{G}^{j_{n-1}}_2 \hat{G}^{j_{n-2}}_a \cdots
\hat{G}^{j_3}_2\hat{G}^{j_2}_a\hat{G}^{j_1}_1 |s_1\rr=0.\label{prop}
\end{equation}

Using the matrix description given above we can calculate explicitly a
power of auxiliary search $\hat G_a$:
\begin{equation}\begin{array}{c}
\hat G_a^{j_a}= \\ {}\\ \left(\begin{array}{ccc} c(2j_a \theta_2)&-
c(2\gamma)s(2j_a\theta_2) & s(2\gamma)s(2j_a\theta_2) \\
c(2\gamma)s(2j_a\theta_2) & s^2(2\gamma)
+c^2(2\gamma)c(2j_a\theta_2) &
s(2\gamma)c(2\gamma)[1-c(2j_a\theta_2)]\\
-s(2\gamma)s(2j_a\theta_2)\; &
s(2\gamma)c(2\gamma)[1-c(2j_a\theta_2)]\; &
c^2(2\gamma)+s^2(2\gamma)c(2j_a\theta_2) \end{array}\right)
\end{array}\end{equation}

Its spectrum is 
\begin{eqnarray}
\hat{G}^{j_2}_a|u_2^{\pm}\rangle =\exp (\pm 2i\theta_2
j_2)|u_2^{\pm}\rangle ,\quad \hat{G}^{j_2}_2|u_2^{0}\rangle
=|u_2^{0}\rangle
\end{eqnarray}
Where the eigenvectors can be written as
\begin{equation}
|u_2^{\pm}\rangle=\frac{1}{\sqrt{2}}\left(\begin{array}{c}
1\\
\mp i \cos (2\gamma) \\
\pm i \sin (2\gamma)
\end{array} \right), \qquad |u_2^{0}\rangle= \left(\begin{array}{c}
0\\
\sin (2\gamma) \\
\cos (2\gamma)
\end{array} \right).
\end{equation}
To summaries in this section: we eliminated reflection by introducing a new 
element $\hat{G}a$. In new formulation of the GRK algorithm (see (\ref{prop})
only even powers of $\hat{G}_1$ appears. These means that in formulae 
(\ref{reflection}) and (\ref{mat})
$j_1$ can be replaced by  even number, so $(-1)^{j_1}=1$.

\section{Lie Algebra Relations}
The introduction of a third operator $\hat{G}_a$ simplifies the
analysis of general algorithms, since now we do not have to
take into account  reflections.  Now the algorithm consists of a 
 sequence of even powers of  $\hat{G}_1$ and integer powers of  $\hat{G}_2$ and $\hat{G}_a$.
Each of these operators [searches ]
is  an element of  $SO(3)$. This is a  simplification, since now we
are dealing with connected component to the identity element. But
$\hat{G}_a$ is dependent on the other two operators (\ref{ga}).

Any element of the $SU(2)$ group can be written
in terms of rotations around two linearly independent unite vectors
(see the Appendix). This is also true for   $SO(3)$  group.
 To find general relations among our three
operators it is useful to first see what is the Lie Algebra relation.

Using the matrix form of $\hat{G}^{2j}_1$ given by (\ref{mat})  we can compute
its expression for small powers:

\begin{equation}\label{liegone}\hat{G}^{2j}_1= I+ 4\theta_2j{\bf T}_{G_1}= I+4\theta_2 j \left(
\begin{array}{ccc} 0&\sin^2{(\gamma)} & \sin{(\gamma)}\cos{(\gamma)} \\
- \sin^2{(\gamma)} & 0&0 \\
-\sin{\gamma}\cos{\gamma}&0&0\end{array}\right)
\end{equation}

The same calculation can be done with $\hat{G}_2$

\begin{equation}\label{liegtwo}\hat{G}^{j}_2= I+ 2\theta_2j{\bf T}_{G_2}= I+2\theta_2 j \left(
\begin{array}{ccc} 0&1 &0 \\
- 1 & 0&0 \\
0&0&0\end{array}\right),
\end{equation}
and $\hat{G}_a$
\begin{equation}\label{liega}\hat{G}^{j}_a= I+ 2\theta_2j{\bf T}_{G_a}= I+2\theta_2j \left(
\begin{array}{ccc} 0&-\cos{(2\gamma)} & \sin{(2\gamma)} \\
\cos{(2\gamma)} & 0&0 \\
-\sin{2\gamma}&0&0\end{array}\right),
\end{equation}
note that we used the relation  $\theta_1= \sin (\gamma )\  \theta_2$
to simplify the above equations. The relation follows from the definition 
of the angles in the limit of $b\rightarrow \infty $. 

 ${\bf T}_{G_1}$, ${\bf T}_{G_2}$ and ${\bf T}_{G_2}$ are 
as elements of  Lie Algebra generators corresponding to our searches.
Using their matrix expressions we see that they satisfy the linear
relation
{\begin{equation}
{\bf T}_{G_a}+ {\bf T}_{G_2}-2{\bf T}_{G_1}=0 \label{lie}
\end{equation}
which explicitly shows that the Lie algebra elements describing global, local and auxiliary  searches
 are linearly dependent.
In the next section we shall rise this relation into the group, see (\ref{group}).

\section{$SU(2)$ Formulation}
\label{sutwo}

Let us formulate partial search in terms of 
 $su(2)$ algebra and later
$SU(2)$ group. The transition to $SU(2)$ group makes the
manipulation of the group elements algebraically easier.

From Eqs. \ref{liegone}, \ref{liegtwo} and \ref{liega} we see that
the Lie algebra generators are linear combinations of standard
generators $T_z$ and $T_y$ of the $so(3)$ Lie algebra, see (\ref{gener}). Any three
dimensional vector (\ref{vector}) can be mapped to two dimensional matrices 
$$V\to v=a\sigma_x+b\sigma_y+c\sigma_z,$$
using Pauli matrices 
\begin{equation}\sigma_x=\left(\begin{array}{cc}0&1\\1&0\end{array}\right),
\quad \sigma_y=\left(\begin{array}{cc}0&-i\\i&0\end{array}\right),\quad
\sigma_z=\left(\begin{array}{cc}1&0\\0&-1\end{array}\right).
\end{equation}

We can map $so(3)$ algebra to $su(2)$ algebra by  replacing the generators $T_x$, $T_y$ and $T_z$
\begin{equation}\label{gener}
T_x= \left(
\begin{array}{ccc} 0& 0 &0 \\
0 & 0&1 \\
0&-1&0\end{array}\right),\\
\qquad 
T_y= \left(
\begin{array}{ccc} 0&0 &-1 \\
0 & 0&0 \\
1&0&0\end{array}\right),\\
\qquad
T_z= \left(
\begin{array}{ccc} 0&1 &0 \\
- 1 & 0&0 \\
0&0&0\end{array}\right),
\end{equation}

by $\frac{i}{2}\sigma_x$, $\frac{i}{2}\sigma_y$ and $\frac{i}{2}\sigma_z$ correspondingly,so
$$R=e^{a\cdot T}\to {\bf u}=e^{\frac{i}{2}a\cdot \sigma}, $$
and rotation act on vectors, 
$$RV \to {\bf u}v{\bf u}^{-1}$$
The actual $SU(2)$ to  $O(3)$ correspondence is given by
$$
{\bf u} (\vec r\cdot \vec{\sigma}){\bf u}^{-1}= (R \vec{r})\cdot \vec{\sigma}
$$
Here $\vec{r}$ is a vector with real components.
Elements of partial search can be mapped to $SU(2)$ group in the following way:

$$\hat G_1^{j_1}\to {\bf u}^{j_1}_{1}=\left(\begin{array}{cc}\cos(j_1\theta_1)+
i\sin(\gamma)\sin(j_1\theta_1)&-\cos(\gamma)\sin(j_1\theta_1)\\
\cos(\gamma)\sin(j_1\theta_1)& \cos(j_1\theta_1)-
i\sin(\gamma)\sin(j_1\theta_1) \end{array}\right)$$
$$\hat G_2^{j_2}\to {\bf u}_{2}=\left(\begin{array}{cc}e^{i(j_2\theta_2}&0\\0&
e^{-i(j_2\theta_2)}\end{array}\right)$$
$$\hat G_a^{j_a}\to {\bf u}^{j_a}_{a}=\left(\begin{array}{cc}\cos(j_a\theta_2)-
i\cos(2\gamma)\sin(j_a\theta_2)&-\sin(2\gamma)\sin(j_a\theta_2)\\
\sin(2\gamma)\sin(j_a\theta_2)&\cos(j_a\theta_2)+
i\cos(2\gamma)\sin(j_a\theta_2) \end{array}\right)$$

So we mapped partial search in $SU(2)$ group.
 The elements ${\bf u}_1$, ${\bf u}_2$ and ${\bf u}_a$ of $SU(2)$ group  are  dependent.
Using the Appendix we  found algebraic relation between group elements
describing global, local and auxiliary searches:
\begin{equation}
{\bf u}_1^{j_1}{\bf u}_a^{-j_2}{\bf u}_1^{j_1}={\bf u}_2^{j_2},\label{group}
\end{equation}
here $\sin(\gamma)\tan(j_2\theta_2)=\tan(j_1\theta_1)$.
Corresponding Lie algebraic relation is (\ref{lie}).

\section{Conclusion}

In this paper we formulated the  partial
search algorithms in terms of group theory.
We think that it will be useful for proof of optimality of GRK algorithm
in wide class of partial search algorithms.

\section{Acknowledgment}
Our work was funded by NSF grant DMS-0503712.

\section{Appendix}
Arbitrary element of 
$SU(2)$ group can be written using only rotation around   two different 
axis, see page 176 of the book \cite{theorem}. We shall represent rotation axis by a unit vector.
If we define  $R_{\vec{n}}(\lambda)$ as a rotation around the unit vector
$\vec{n}$ by an angle $\lambda$:
$$ R_{\vec{n}}(\lambda)=\exp\{-i\frac{\lambda}{2}(\vec{n}\cdot \vec{\sigma}) \} $$
 A  rotation around any axis by any angle  (arbitrary element of $SU(2)$ group $R$) can 
be represented  as sequential   rotations around two fixed axis  $\vec{n}$ and $\vec{m}$:
$$R=R_{\vec{n}}(\lambda)R_{\vec{m}}(\theta)R_{\vec{n}}(\gamma),$$
Here $\vec{n}$ and $\vec{m}$ are two linearly independent unit
vectors and $(\lambda,\theta,\gamma)$ are three real numbers (angles).

\end{document}